# Afterglow Linear Polarization Signatures from Shallow GRB Jets: Implications for Energetic GRBs

Gal Birenbaum[1], Ramandeep Gill[2,3], Omer Bromberg[1], Paz Beniamini[3,4,5], and Jonathan Granot[3,4,5]
[1] Tel Aviv University, The Raymond and Beverly Sackler School of Physics and Astronomy, Tel Aviv University, Tel Aviv 69978, Israel
[2] Instituto de Radioastronomía y Astrofísica, Universidad Nacional Autónoma de México, Antigua Carretera a Pátzcuaro # 8701, Ex-Hda. San José de la Huerta, Morelia, Michoacán, México C.P. 58089, Mexico
[3] Astrophysics Research Center of the Open University (ARCO), The Open University of Israel, P.O. Box 808, Ra'anana 4353701, Israel
[4] Department of Natural Sciences, The Open University of Israel, P.O. Box 808, Ra'anana 4353701, Israel
[5] Department of Physics, The George Washington University, Washington, DC 20052, USA
Received 2024 June 8; revised 2024 August 15; accepted 2024 August 16; published 2024 October 18

## Abstract

Gamma-ray bursts (GRBs) are powered by ultrarelativistic jets. The launching sites of these jets are surrounded by dense media, which the jets must cross before they can accelerate and release high-energy emission. Interaction with the medium leads to the formation of a mildly relativistic sheath around the jet, resulting in an angular structure to the jet's asymptotic Lorentz factor and energy per solid angle, which modifies the afterglow emission. We build a semi-analytical tool to analyze the afterglow light curve and polarization signatures of jets observed from a wide range of viewing angles, and focus on ones with slowly declining energy profiles known as shallow jets. We find overall lower polarization compared to the classical top hat jet model. We provide an analytical expression for the peak polarization degree as a function of the energy profile power-law index, magnetic field configuration, and viewing angle, and show that it occurs near the light-curve break time for all viewers. When applying our tool to GRB 221009A, suspected to originate from a shallow jet, we find that the suggested jet structures for this event agree with the upper limits placed on the afterglow polarization in the optical and X-ray bands. We also find that at early times the polarization levels may be significantly higher, allowing for a potential distinction between different jet structure models and possibly constraining the magnetization in both forward and reverse shocks at that stage.

*Unified Astronomy Thesaurus concepts:* Gamma-ray bursts (629); Non-thermal radiation sources (1119); Transient sources (1851); Relativistic jets (1390)

## 1. Introduction

The understanding that gamma-ray bursts (GRBs) originate from narrow, ultrarelativistic jets is based on three main observational lines of evidence. (i) The high observed power in prompt gamma rays requires the flow to move with a high Lorentz factor to reduce the pair production and Thomson scattering opacity and allow the gamma-ray photons to escape without substantially changing the spectrum (J. H. Krolik & E. A. Pier 1991; E. Woods & A. Loeb 1995; Y. Lithwick & R. Sari 2001). (ii) The high observed fluence, which calls for narrow emission sites to avoid an energy crisis in emission from a spherical flow (T. Piran 1999). (iii) Achromatic "jet breaks" in the afterglow light curves, which occur when the Lorentz factor of the decelerating flow becomes comparable to the inverse of the jet half-opening angle, making the entire jet visible and leading to a steepening of the light curve at later times (J. E. Rhoads 1997, 1999; R. Sari et al. 1999). Typically, the observed afterglow light curve has a power-law behavior in frequency ($\nu$) and time ($t_{\rm obs}$), such that $F_\nu \propto \nu^{-\alpha} t_{\rm obs}^{-\beta}$. The simple model of a sharp-edged jet, with uniform energy and Lorentz factor distributions and no sideways expansion (known as a "top hat" jet), produces light curves that should steepen by $\Delta\beta \gtrsim \frac{1}{2}$ during the jet break. The change in the temporal index, as well as the duration of the jet break, depend on the density profile of the external medium. Such steepenings were seen in about half of Swift long GRBs (A. Panaitescu 2007; X.-G. Wang et al. 2018) and were consistent with an average jet half-opening angle of $\sim 6.5°$, confirming this picture (J. L. Racusin et al. 2009).

The traditional view of this sharp-edged, "top hat" jet was challenged with the landmark discovery of GRB 170817A/GW170817. Analysis of the afterglow radio light curve showed that the jet was observed most likely at a viewing angle $\sim 3-6$ times larger than the inferred half-opening angle of the relativistic jet (R. Gill & J. Granot 2018; D. Lazzati et al. 2018; K. P. Mooley et al. 2018, 2022; G. Ghirlanda et al. 2019; R. Gill et al. 2019; E. Troja et al. 2019; T. Govreen-Segal & E. Nakar 2023). Such an observation is possible if the jet's energy per solid angle and Lorentz factor gradually and continuously drop with the angle $\theta$ from its symmetry axis outside of some core angle $\theta_c$, with part of the outflowing material directed along the line of sight (LOS). In particular, the jet core may be surrounded by a wide-angle mildly relativistic outflow component. Such structure was seen in several numerical simulations of GRB jets breaking out of the confining medium of the progenitor system.[6] It originates from instability processes on the jet boundary, such as the Rayleigh–Taylor instability (J. Matsumoto et al. 2017; J. Matsumoto & Y. Masada 2019; O. Gottlieb et al. 2021; P. Abolmasov & O. Bromberg 2023), which mix jet with medium material and typically has a power-law angular profile in energy density and Lorentz factor. The degree of mixing and amount of energy in the mildly relativistic component depends on the properties of both the jet and confining medium, and contains information on the conditions at the jet injection site (e.g., O. Gottlieb et al. 2021).

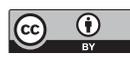

---

[6] The envelope of a collapsing star in the case of long GRBs or the subrelativistic neutron star merger ejecta in the case of short GRBs.

In several cases of powerful GRBs with long-lasting X-ray afterglows and very high-energy emission, the afterglow light curves show a gradual transition in their temporal decay indexes during the jet break with late time values of $1.4 \lesssim \beta \lesssim 1.7$, rather than a relatively sharp steepening to a canonical value of $\beta \sim 2.2$ (B. O'Connor et al. 2023). Such a behavior fits a case where the energy in the mildly relativistic component is comparable to the energy in the relativistic core, leading to a shallow angular distribution of the energy per solid angle and Lorentz factor in the jet. An example of such a case is the extremely luminous GRB 221009A, which showed only a slight steepening in its X-ray afterglow. The multiwavelength afterglow light curves were fitted with a narrow cored jet surrounded by a shallow structure in both the energy per unit solid angle and Lorentz factor distributions with combined contributions from the reverse and forward shocks (R. Gill & J. Granot 2023). The fit, however, is not unique and other similar shallow jet models are able to describe the afterglow light curves (e.g., B. O'Connor et al. 2023). To determine the jet structure, it is important to be able to distinguish between the different jet models that fit the light curve.

In addition to the observed light curve, upper limits were placed on the linear polarization of GRB 221009A by M. Negro et al. (2023) 3.5 days post-burst at 13.8% and 8.3% in the X-ray and optical bands, respectively. This is the first time limits on the X-ray afterglow polarization were obtained for any GRB and in some cases, theory predicts X-ray polarization levels would be higher than optical ones (J. Granot 2003; E. M. Rossi et al. 2004; G. Birenbaum & O. Bromberg 2021; J. Shimoda & K. Toma 2021). This observation was made using the Imaging X-ray Polarization Explorer (IXPE) instrument in the 2–8 keV band (M. Negro et al. 2023).

Unlike the total flux, linear polarization is sensitive to the angular distribution of energy and emissivity across the jet, making it a promising way to distinguish between different jet structures. E. M. Rossi et al. (2004) explored the temporal behavior of linear polarization signatures from jets with uniform (top hat), Gaussian and a power-law angular energy distribution that follows $\theta^{-2}$ from various viewing angles, discussing their overall differences. R. Gill & J. Granot (2018) fitted different jet structures to the X-ray and radio afterglow light curves of GW170817 and modeled the consequent temporal evolution of the radio afterglow polarization. In a follow-up work, R. Gill & J. Granot (2020) used the structures they found, combined with an upper limit on the radio afterglow polarization of GW170817, and obtained a limit on the magnetic field structure behind the forward shock. This is also done independently by A. Corsi et al. (2018), E. Stringer & D. Lazzati (2020), and O. Teboul & N. J. Shaviv (2021).

In this work we look at jets with power-law energy and Lorentz factor distributions in their wings, motivated by results from numerical simulations (O. Gottlieb et al. 2021). We construct a semi-analytical tool that explores the evolution of the afterglow light curve and linear polarization of the forward shock for various power-law profiles, magnetic field configurations behind the forward shock, and observer viewing angle. We focus on the difference in the observational signature of jets with shallow and steep angular profiles. The semi-analytical approach of the tool we outline in this work is similar to that of the public code afterglowpy, which calculates afterglow light curves analytically in order to fit observed data (G. Ryan et al. 2020). However, our tool is able to account for the forward shock evolution as it plunges into power-law external medium density profiles. We refer the reader to Appendix F for further comparisons between publicly available afterglow codes and our own.

We obtain an analytical expression for the dependence of the peak polarization degree on the various system parameters. Finally, we employ our method on the observations of GRB 221009A and fit two different shallow structures (R. Gill & J. Granot 2023; B. O'Connor et al. 2023) to the optical and X-ray light curves. We find the produced polarization curves agree with the upper limits on the linear polarization in both energy bands (M. Negro et al. 2023).

## 2. Methods

To model the polarized afterglow emission from the forward shock we expand the numerical method developed in G. Birenbaum & O. Bromberg (2021). We model a relativistic shock decelerating in a cold ambient medium with a power-law rest-mass density profile, $\rho \propto R^{-k}$. The emitting region is approximated as a 2D surface located just behind the shock radiating synchrotron emission. To calculate the emission, we divide the surface into small cells, where in each cell we place a uniform magnetic field with some 3D orientation. The comoving magnetic field direction, $\hat{\boldsymbol{B}}'$, and its corresponding absolute value, $B'$, are randomly drawn from a probability distribution corresponding to an isotropic field (in 3D) that is stretched in the radial direction by a factor of $\xi$ (as described in Appendix B). We then calculate the $I_\nu$, $Q_\nu$, $U_\nu$ Stokes parameters in each cell and sum up the contributions from all cells to calculate the total emission and polarization. Below we briefly describe the methods we use for each step of the calculation. A more detailed description can be found in Appendixes A–C and in G. Birenbaum & O. Bromberg (2021).

### 2.1. Jet Structure and Forward Shock Kinematics

We assume a conical jet propagating in a medium with a power-law density profile $\rho(R) = m_p n(R) = AR^{-k}$, where $m_p$ is the proton mass, $n$ is the number density of the ambient medium, and $A = m_p n(R_0) R_0^k$ is evaluated at a reference radius $R_0$. The propagation leads to the formation of a semispherical shock with the same opening angle as the jet, which emits the afterglow radiation. The jet has an angular structure consisting of a uniform core surrounded by extended wings with power-law distributions of kinetic energy and Lorentz factor (R. Gill & J. Granot 2018). The structure is expressed in terms of the angular profile function:

$$\Theta \equiv \sqrt{1 + \left(\frac{\theta}{\theta_c}\right)^2}, \qquad (1)$$

where $\theta$ is the polar angle measured from the jet symmetry axis and $\theta_c$ is the half-opening angle of the jet core. We define the profiles of the energy per solid angle, $\epsilon = dE/d\Omega$, and the initial

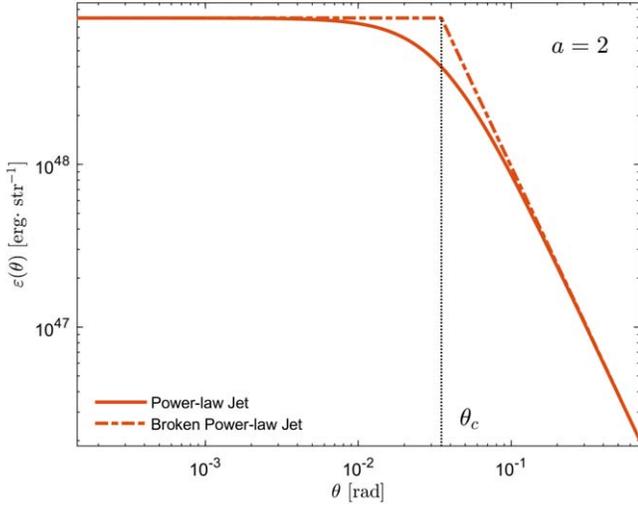

**Figure 1.** Two of the angular energy profiles considered in this work, featuring a core opening angle of $\theta_c = 2°$ and power-law wings with $a = 2$. We study the broken power-law jet (red dashed–dotted line) in the context of on-axis structured jets in order to focus on the interplay between the core and the jet wings and corresponding edge effects. The smooth power-law jet is investigated with off-axis jets (red solid line).

(coasting phase) Lorentz factor, $\Gamma_0$, according to

$$\epsilon(\theta) = \epsilon_c \Theta^{-a}, \quad (2)$$

$$\Gamma_0(\theta) = 1 + (\Gamma_c - 1)\Theta^{-b}, \quad (3)$$

where $\epsilon_c$ and $\Gamma_c$ are the jet core energy per unit solid angle and the initial Lorentz factor of the shocked radiating fluid behind the forward shock, respectively. This structure allows for a smooth transition between the core and wings. We name this structure a "power-law jet" (see Figure 1, solid line) and use it to explore the emission seen by observers located at angles larger than $\theta_c$ (off-axis observers). Another type of structure we investigate is the "broken power-law jet," which we explore in the context of observers located at an observing angle $\theta_{obs} < \theta_c$ (on-axis observers). In such jets, it is easier to control $\theta_c$ when changing the power-law structure of the wings. (see Figure 1, dashed–dotted line). The structure of the broken power-law jet is defined as

$$\epsilon(\theta) = \begin{cases} \epsilon_c & \theta \leqslant \theta_c, \\ \epsilon_c \left(\dfrac{\theta}{\theta_c}\right)^{-a} & \theta > \theta_c. \end{cases} \quad (4)$$

We follow the propagation of the forward shock from the coasting phase through the relativistic and nonrelativistic deceleration phases. To model the shock dynamics, we assume that each cell in the shock at an angle $\theta$, propagates as if it was part of a spherical shock with an energy $E_{k,iso} = 4\pi\varepsilon(\theta)$ and an initial Lorentz factor $\Gamma_0(\theta) = [1 - \beta_0^2(\theta)]^{-1/2}$ (R. Gill & J. Granot 2018; P. Beniamini et al. 2020). The dynamics of each cell are scaled according to the deceleration radius

$$R_{dec}(\theta) = \left[\frac{(3-k)E_{k,iso}(\theta)}{4\pi A c^2 \Gamma_0^2(\theta)\beta_0^2(\theta)}\right]^{1/(3-k)}, \quad (5)$$

where $c$ is the speed of light. At this radius the shock begins to decelerate according to the Blandford–Mckee solution (R. D. Blandford & C. F. McKee 1976; R. Gill & J. Granot 2018), for as long as it remains relativistic. The Lorentz factor evolution of each cell can be described in terms of the normalized shock radius at the cell location, $\zeta \equiv R/R_{dec}$, and the initial Lorentz factor $\Gamma_0$,

$$\Gamma(\zeta, \Gamma_0)$$
$$= \frac{\Gamma_0 + 1}{2 \zeta^{3-k}}\left[\sqrt{1 + \frac{4\Gamma_0}{\Gamma_0 + 1}\zeta^{3-k} + \left(\frac{2\zeta^{3-k}}{\Gamma_0 + 1}\right)^2} - 1\right]. \quad (6)$$

This relation follows the hydrodynamic evolution of each cell from the coasting phase, through the Blandford–Mckee phase, until it becomes nonrelativistic and converges to the Sedov–Taylor solution (G. Taylor 1950; L. I. Sedov 1959).

### 2.2. The Emission and the Emitting Surface

The emission is modeled as slow-cooling synchrotron radiation, originating from electrons located on a 2D surface at the shock immediate downstream that has a bulk Lorentz factor[7] $\Gamma$. The properties of the emitting electrons, such as the downstream density $n' = 4\Gamma n$, are obtained from the shock jump conditions. We use a Taub–Matthews adiabatic index[8] in the equation of state, which transitions smoothly from $4/3$ to $5/3$ when $\Gamma \approx 1$. We further assume that a fraction $\chi_e$ of the electrons are accelerated to relativistic velocities behind the shock and emit the observed synchrotron radiation (e.g., D. Eichler & E. Waxman 2005; G. Ryan et al. 2020). The emission spectrum in each cell is calculated in the shocked fluid frame, with critical frequencies and a peak spectral power (per unit frequency and per unit volume) (R. Gill & J. Granot 2018)

$$\nu'_m = \frac{3}{\sqrt{2\pi}}\left(\frac{p-2}{p-1}\right)^2 \frac{q_e}{m_e^3 c^5} S(\xi)\sqrt{\epsilon_B}\,\epsilon_e^2 e'^{5/2}(\chi_e n')^{-2} \sin\psi', \quad (7)$$

$$\nu'_c = \frac{27}{32}\sqrt{\frac{2}{\pi}}\frac{q_e m_e c}{\sigma_t^2}(S(\xi))^{-3}(\epsilon_B e')^{-3/2}\left(\frac{\Gamma}{t}\right)^2 \sin\psi', \quad (8)$$

$$P'_{\nu',max} = 0.88\frac{1024}{27}\sqrt{\frac{8}{\pi}}\frac{p-1}{3p-1}\frac{q_e^3}{m_e c^2}$$
$$S(\xi)\sqrt{\epsilon_B e'}\,(\chi_e n')\sin\psi', \quad (9)$$

where $t$ is the lab frame time, $\sigma_t$ is the Thomson cross section, $m_e$ is the electron mass, and $q_e$ is its charge. In addition, $p$ is the power-law index of the energy distribution of the radiating electrons with $N(\gamma_e) \propto \gamma_e^{-p}$ for $\gamma_e > \gamma_m$, $\epsilon_e$ and $\epsilon_B$ are the energy fractions that go to the electrons and the magnetic field, respectively, $S(\xi)$ is the stretching factor of the magnetic field according to the choice of $\xi$, the magnetic field structure parameter (Equation (B4)), $e'$ is the bulk internal energy density

$$e' = (\Gamma - 1)n'm_p c^2, \quad (10)$$

and $\psi'$ is the pitch angle of the electrons gyrating around the magnetic field lines. Since the electrons are highly relativistic,

---
[7] The actual Lorentz factor of the shock is $\sqrt{2}\,\Gamma$ where $\Gamma$ is the Lorentz factor of its immediate downstream, however, we neglect this factor in our calculation following R. Gill & J. Granot (2018).
[8] $\hat{\gamma} = \frac{4\Gamma + 1}{3\Gamma}$.

their emission is beamed into a cone of angular size $1/\gamma_e$ around their velocity vectors and the observer sees emission from electrons at a pitch angle $\psi' = \arccos(\hat{\boldsymbol{B}}' \cdot \hat{\boldsymbol{n}}'_{\rm obs})$, where $\hat{\boldsymbol{n}}'_{\rm obs}$ and $\hat{\boldsymbol{B}}'$ are unit vectors pointing toward the observer and along the direction of the magnetic field in the proper frame respectively (see Equations (A1), (C3)). The corresponding spectral power per unit volume in the slow-cooling regime, which is typically valid during most of the afterglow phase, maintains

$$P'_{\nu'} = P'_{\nu',\max} \begin{cases} \left(\dfrac{\nu'}{\nu'_{\rm m}}\right)^{1/3} & \nu' < \nu'_{\rm m} \\ \left(\dfrac{\nu'}{\nu'_{\rm m}}\right)^{-(p-1)/2} & \nu'_{\rm m} < \nu' < \nu'_{\rm c} \quad (11) \\ \left(\dfrac{\nu'_{\rm c}}{\nu'_{\rm m}}\right)^{-(p-1)/2}\left(\dfrac{\nu'}{\nu'_{\rm c}}\right)^{-p/2} & \nu'_{\rm m} < \nu'_{\rm c} < \nu'. \end{cases}$$

From this we calculate the proper isotropic-equivalent luminosity by multiplying the spectral power per unit volume (Equation (11)) by the emitting volume

$$L'_{\rm iso, \nu'} = P'_{\nu'} \cdot 4\pi R^2 \Delta', \quad (12)$$

with $\Delta' \equiv \dfrac{R}{4(3-k)\Gamma}$ (R. Gill & J. Granot 2018).

To calculate the total observed emission we sum the contributions from all cells that have the same photon arrival time to the observer. These cells constitute a surface known as the equal arrival time surface (EATS; R. Sari 1997; J. Granot et al. 1999). To associate the EATS radius at each angle with an arrival time to the observer $t_{\rm obs}$, we numerically solve the equation (see Appendix C)

$$\dfrac{t_{\rm obs}}{(1+z)} - \int_0^R \left(\dfrac{1 - \beta\tilde{\mu}}{\beta}\right)\dfrac{dR}{c} = 0, \quad (13)$$

where $\tilde{\mu} = \hat{\boldsymbol{\beta}} \cdot \hat{\boldsymbol{n}}'_{\rm obs}$. We then employ Equation (6) to calculate the corresponding $\Gamma(R)$ and use both values to calculate the emission observed at a time $t_{\rm obs}$. The overall observed flux is found by integrating over the emission from all cells on the EATS and accounting for the source redshift, $z$, giving

$$F_\nu(t_{\rm obs}) = \dfrac{(1+z)}{16\pi^2 d_{\rm L}^2} \int D^3(\tilde{\mu}) L'_{\rm iso, \nu'} d\Omega, \quad (14)$$

where $d_{\rm L}$ is the luminosity distance, $\tilde{\mu} = \hat{\boldsymbol{\beta}} \cdot \hat{n}_{\rm obs}$ with $\hat{\boldsymbol{\beta}}$ as the unit vector in the velocity direction of the emitting cell, and $D(\tilde{\mu}) = [\Gamma(1 - \beta\tilde{\mu})]^{-1} = (1+z)\nu/\nu'$ is the Doppler factor.

### 2.3. The Observed Polarization

Since the emission in each cell originates from a uniform field it is linearly polarized with a polarization degree of

$$\Pi_{\max} = \dfrac{\alpha + 1}{\alpha + 5/3}, \quad (15)$$

where $\alpha$ is the spectral index of the emission at the given frequency, $P'_{\nu'} \propto \nu'^{-\alpha}$, evaluated according to Equation (11). The local direction of the polarization vector in the proper frame is

$$\hat{\boldsymbol{P}}' = \dfrac{\hat{\boldsymbol{n}}'_{\rm obs} \times \hat{\boldsymbol{B}}'}{|\hat{\boldsymbol{n}}'_{\rm obs} \times \hat{\boldsymbol{B}}'|}. \quad (16)$$

This quantity is transformed to the observer frame and projected onto the plane of the sky according to the procedure described in Appendix D. The polarization angle in each cell is then calculated from the ratio of the two orthogonal polarization vector components in the observer frame:

$$\phi_{\rm p,0} = \arctan\left(\dfrac{P_{\rm y}}{P_{\rm x}}\right), \quad (17)$$

where $\hat{x}$ points along the vector connecting the symmetry axis with the LOS on the plane of the sky (for a choice of $\phi_{\rm obs} = 0°$) and $\hat{y}$ points along the orthogonal direction (see Figure 10 for a detailed illustration). We then use Equations (12), (14), and (17) to calculate the flux weighted Stokes parameters at each cell and sum them up over all cells:

$$\dfrac{Q_\nu}{I_\nu} = \dfrac{\int \Pi_{\max} \cos(2\phi_{\rm p,0}) D^3(\tilde{\mu}) L'_{\rm iso, \nu'} d\Omega}{\int D^3(\tilde{\mu}) L'_{\rm iso, \nu'} d\Omega},$$

$$\dfrac{U_\nu}{I_\nu} = \dfrac{\int \Pi_{\max} \sin(2\phi_{\rm p,0}) D^3(\tilde{\mu}) L'_{\rm iso, \nu'} d\Omega}{\int D^3(\tilde{\theta}) L'_{\rm iso, \nu'} d\Omega}. \quad (18)$$

Using the integrated Stokes parameters from Equation (18), we obtain the total observed polarization degree and polarization angle:

$$P_{\rm tot} = \dfrac{\sqrt{Q_\nu^2 + U_\nu^2}}{I_\nu} = \dfrac{Q_\nu}{I_\nu}, \quad (19)$$

$$\phi_{\rm p} = \dfrac{1}{2} \arctan\left(\dfrac{U_\nu}{Q_\nu}\right), \quad (20)$$

where due to the axial symmetry of the flow, $U_\nu$ vanishes, leaving us with $P_{\rm tot} = Q_\nu/I_\nu$ for a choice of $\phi_{\rm obs} = 0°$. This axial symmetry limits changes in the polarization angle to rotations by $90°$ only. The sign of the observed polarization degree changes when such a rotation occurs, as shown in Section 3.1.1.

The time variation of the overall polarization degree and the polarization angle depends on the jet structure and on the magnetic field configuration. It also depends on $\theta_{\rm obs}$ and on the observed frequency, which together with the afterglow parameters, determine the relevant power-law segment (PLS) in the synchrotron spectrum (R. Sari et al. 1998; J. Granot & R. Sari 2002). Equation (18) gives a higher weight to more luminous areas, causing them to dominate the observed polarization. If the emission is dominated by a small area over which the local polarization vector is ordered, the observed polarization would be extremely high (for example, when emission from structured jets turns core dominated at late times). If the emission originates from an extended region in the shock where the direction of the polarization vector varies over a range of $\sim 90°$, some cancellations will arise when integrating the polarization, leading to a moderate to low polarization degree (at early times, the emission from off-axis structured jets is dominated by their extended wings).

**Table 1**
Parameter Space and Afterglow Model for the Cases Considered in Section 3

| Parameter Space | |
|---|---|
| Parameter | Value |
| $\theta_c$ | 2° |
| $\Gamma_c$ | 250 |
| $a$ (PL index of energy structure) | 0.5, 1, 2, top hat jet |
| $b$ (PL index of Lorentz factor structure) | 0 |
| $E_c$ [erg] | $10^{50}$ |
| $n_{\rm ISM}$ [cm$^{-3}$] | 1 |
| Density profile ($n \propto r^{-k}$) | $k = 0$ |
| $\nu_{\rm obs}$ [Hz] | $10^{15}$ (PLS G) |
| $p$ | 2.5 |
| $\epsilon_e$ | 0.1 |
| $\epsilon_B$ | 0.005 |
| $\chi_e$ | 1 |
| $d_L$ [cm] | $10^{28}$ |
| $z$ | 0.54 |

## 3. Results

In this paper, we study the effects of the jet structure and magnetic field 3D orientation on the temporal evolution of polarization seen by viewers from different viewing angles. We focus the study on shallow jets ($a \leqslant 2$) observed in the optical band so that the observed frequency $\nu$ will not cross any critical frequency[9] ($\nu_m < \nu < \nu_c$). The details of the afterglow model we use are given in Table 1. Below we show the results for jets observed from viewing angles within the jet core, with normalized viewing angles $q \equiv \theta_{\rm obs}/\theta_c < 1$ (on-axis jets) and jets observed from viewing angles outside the core, i.e., with $q > 1$ (off-axis jets). All cases considered in this section are set with a flat distribution of the initial Lorentz factor with $b = 0$.

### 3.1. Viewing Angle and Jet Structure

#### 3.1.1. On-axis Jets (q < 1)

Figure 2 shows the temporal evolution of the polarization degree (upper panel) and the afterglow light curve (lower panel) of jets with different power-law structures in energy $a = (0.5, 1, 2)$, compared to a top hat jet, and random magnetic fields confined to the shock surface ($\xi \to 0$), observed at a viewing angle $q = 0.7$. Each structure is marked by a different color, as shown in the figure legend. The polarization degrees of all structures show two peaks at opposing signs corresponding to polarization angles differing by 90°. This polarization signature is well known in the context of top hat jets and is formed due to the fact that the observed emission comes from a ring-like region with an angular radius of $\frac{1}{\Gamma}$ centered around the observer LOS, rather then the entire shock surface (see Figure 6 for an illustration of this ring).[10] In the case of a random magnetic field, confined to the plane of the forward shock, the polarization vector is radially ordered along this ring (on the plane of the sky, relative to the center of the ring-like image). At early times, the entire ring is visible and the polarization degree is zero. As the shock decelerates, the angular radius of the ring grows, and eventually, parts of the polarized ring disappear beyond the jet edge, reducing the cancellations of opposing polarization directions and giving rise to a global polarization degree (see Figure 6 of G. Birenbaum & O. Bromberg 2021 or Figure 3 of R. Sari 1999 for a visual demonstration of this effect). The sign of the polarization degree is determined by the polarization angle where a change in sign means the global polarization vector has rotated by 90°. This rotation occurs when half of the polarized ring disappears beyond the jet edge, followed by a rise in polarization in the opposite direction (G. Ghisellini & D. Lazzati 1999; R. Sari 1999; J. Granot & A. Konigl 2003; L. Nava et al. 2016; G. Birenbaum & O. Bromberg 2021).

---

[9] To see how the polarization degree is affected by spectral breaks, we refer the reader to G. Birenbaum & O. Bromberg (2021), J. Granot (2003), and E. M. Rossi et al. (2004).
[10] A similar polarization signature also occurs for a more uniform (less limb-brightened) afterglow image, as expected in other PLSs, so it is rather robust.

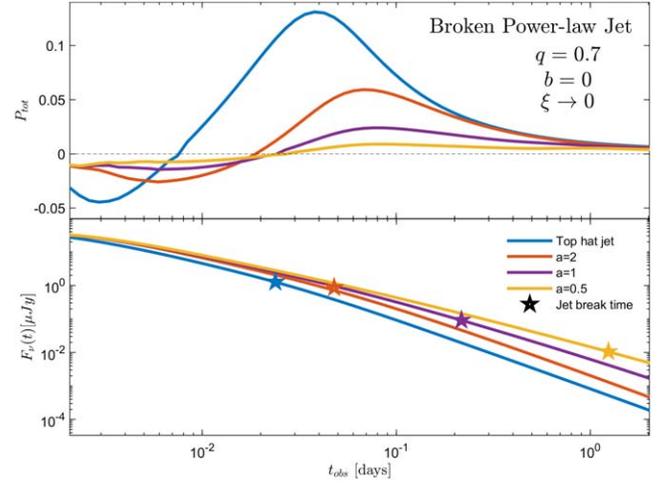

**Figure 2.** Observed polarization degree (upper panel) and flux (lower panel) as a function of time for the broken power-law jet model at observer frequency $\nu = 10^{15}$ Hz (PLS G) with a flat distribution of initial Lorentz factor ($b = 0$) and a random magnetic field structure, confined to the shock plane ($\xi \to 0$). Jet break times for each model are marked with stars. The various values of $a$ represent different slopes for the power-law jet wings and the observer is located at $q = 0.7$. The signs of the polarization degree represent opposing directions of it, with a polarization angle difference of 90°.

When power-law wings are added beyond the uniform core, the effective jet core opening angle becomes larger and the emitting ring reaches the effective core edge at a later time, resulting in the later appearance of the polarization peaks and rotation of the polarization angle. The size of the effective jet core opening angle is dictated by the wing power-law index $a$. In shallower setups (with smaller $a$ values), this effective angle is larger since the contribution to the observed flux from regions at wider angles increases. This effect is shown in Figure 2 with polarization extrema and jet break times, which are marked with stars, occurring at later times for shallower jets. The second polarization peak is associated with the time when the opening angle of the emitting ring is of the order of the jet core opening angle, i.e., when $1/\Gamma \sim \theta_c$. For top hat jets, this is also the typical time when the jet break occurs, resulting in a relatively tight correlation between the two times (R. Sari 1999; J. Granot 2003; J. Granot & A. Konigl 2003; E. M. Rossi et al. 2004; G. Birenbaum & O. Bromberg 2021). This correlation is weaker in shallow jets although it still holds up to a factor 3 between the two times (see Section 3.3 for a detailed discussion). The larger effective jet core opening angle also changes the heights of the polarization peaks as well. This can be explained with the effective normalized observed angle, quantified by the $q$ parameter ($q_{\rm eff}$) becoming smaller. Smaller values of $q_{\rm eff}$ are associated with higher degrees of axial symmetry and thus lower polarization (G. Ghisellini & D. Lazzati 1999; R. Sari 1999; J. Granot & A. Konigl 2003;

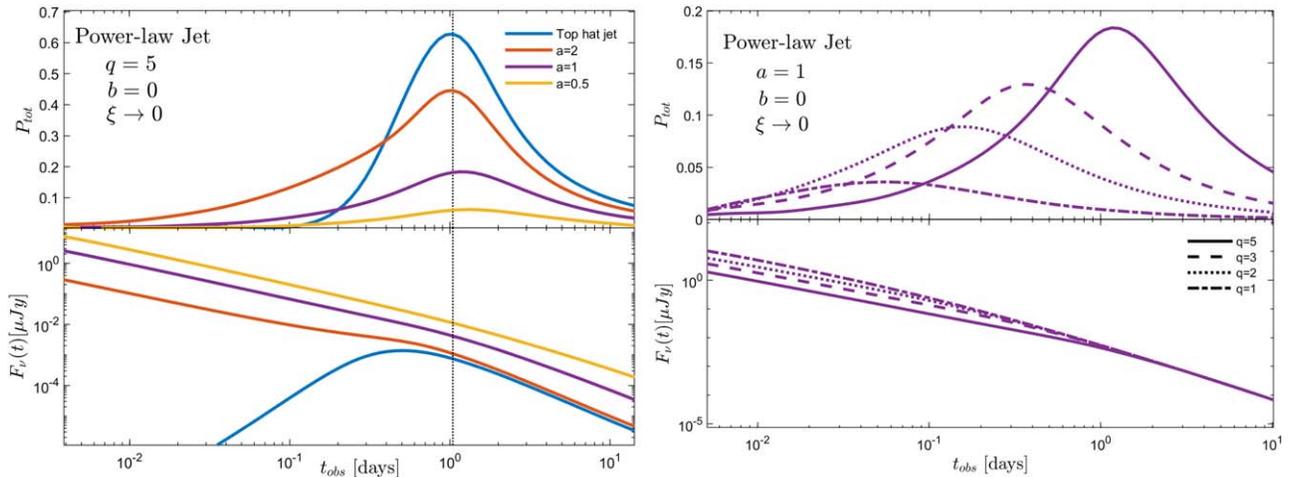

**Figure 3.** Observed polarization degree (upper panels) and flux (lower panels) at observer frequency $\nu = 10^{15}$ Hz (PLS G) with a flat distribution of initial Lorentz factor ($b = 0$) and a random magnetic field structure confined to the face of the shock ($\xi \to 0$) for off-axis jets. The direction of the polarization vector remains constant throughout the temporal evolution. Left panel: we set the viewing angle to be constant for an off-axis jet with $q = 5$ with varying values of $a$. The polarization degree reduces with $a$ and its peak coincides with a break in the light curve. Right panel: the structure of the jet is held constant with $a = 1$ and the off-axis viewing angle is changed. As the observer approaches the jet axis with reducing values of $q$, the polarization peak becomes lower and occurs at earlier times.

E. M. Rossi et al. 2004; G. Birenbaum & O. Bromberg 2021; J. Shimoda & K. Toma 2021; O. Teboul & N. J. Shaviv 2021).

### 3.1.2. Off-axis Jets (q > 1)

In the case of off-axis jets the observer is located at an angle larger than the core half-opening angle. In this case the observed emission is initially dominated by the lower energy per solid angle wing since the core emission is beamed away from the observer due to its high Lorentz factor. As the jet slows down, the beaming relaxes and regions closer to the core gradually become visible, leading to a light-curve evolution that is different than the one in the case of an on-axis jet. While for shallow jets the contribution of the core to the total flux is subdominant, it typically still dominates the polarized flux, and therefore determines the polarization. To study the effect of the jet structure on the time evolution of the polarization signature, we begin by viewing the jet at a viewing angle of $q = 5$, and hold it constant while exploring different power-law profiles of the wings corresponding to shallow jets and comparing them to a top hat jet (see Figure 3, left panel). The configuration of the magnetic field is random on the plane of the shock ($\xi \to 0$). Similarly to the on-axis case, the observed flux comes from a "ring-like" shape centered around the observer's LOS. However, due to the angular structure in energy density and Lorentz factor, the flux distribution in the ring is not uniform but concentrated in a region closer to the jet core. As a result, the observed flux is always polarized in the direction pointing from the LOS toward the jet core, which translates in our setup to a positive sign of $Q_\nu$ (see Figure 6 for a simplified illustration). The peak in the polarization curve occurs when the opening angle of the emitting ring is of order $\theta_{\rm obs}$ and the core beaming cones open toward the observer, revealing it to her with $1/\Gamma \sim \theta_{\rm obs}$ (this time is defined similarly as $t_{\rm sh}$ in P. Beniamini et al. 2022). At this time, the light from the emitting ring reaches maximal asymmetry, maximizing the fraction of polarized light. For this reason, the time of the peak is nearly independent of the energy profile[11] and depends on the viewing angle alone. This can also be seen in the right panel of Figure 3, where the $q$ parameter is changed while keeping the jet structure constant. As in the on-axis case, decreasing the power-law index of the energy structure in the wings reduces the asymmetry in the light distribution and the value of the peak polarization. In the case of a top hat jet (blue curve in Figure 3), only the core is luminous. Within the context of a structured jet, we can say that light is only coming from parts of the emission ring that intersect with the core. During the polarization peak time, the entire core is visible and the emitting region on the ring has an angular size of $\sim 2\theta_c$, which is smaller than the angular size of the ring $2\theta_{\rm obs} \simeq 10\theta_c$ corresponding to the chosen value of $q = 5$ (see Figure 6 for a visual demonstration of this effect). As a result, the radially polarized emission coming from this narrow region in the ring resembles a uniform polarization pattern and the polarization degree reaches a value close to the threshold value given in Equation (15). The continual deceleration of the jet at later times adds light from regions that are progressively further away from the core, and therefore contribute less light. This causes the light curve to show a break similar to the jet break at a time close to the time of the polarization peak. We find that the two times are correlated with each other within a factor of 3 (see upper panel of Figure 5 with further explanations in Section 3.3).

### 3.2. Magnetic Field 3D Orientation

So far we assumed that the magnetic field configuration is random in the plane of the shock with a small coherence length. Although this assumption is often used in describing the magnetic field configuration just behind the shock, it does not take under consideration the effects of turbulence and stretching of the magnetic field further downstream, which can generate a non-negligible radial component (e.g., J. Granot & A. Konigl 2003; R. Gill & J. Granot 2020). Hereon we will denote the comoving field component in the plane of the shock by $B'_\perp$ and its radial component (parallel to the shock normal) by $B'_\parallel$. A parallel field produces local polarization in an azimuthal direction around the LOS, where most of the observed light is coming from a ring with an opening angle $1/\Gamma$ around the LOS (see L. Nava et al. 2016 for a visualization of this effect). This polarization vector orientation is rotated by exactly 90° from the radial polarization produced by a purely perpendicular field (in the shock plane; see Figure 11.8 of J. Granot & E. Ramirez-Ruiz 2012 or G. Birenbaum & O. Bromberg 2021). Since polarization is a pseudo-vector, the combination of the two polarization components lowers the overall polarization.

---
[11] For a fixed value of $\epsilon_c$ there is only a weak dependence on $a$ that arises for sufficiently shallow jets (see P. Beniamini et al. 2022 for details).

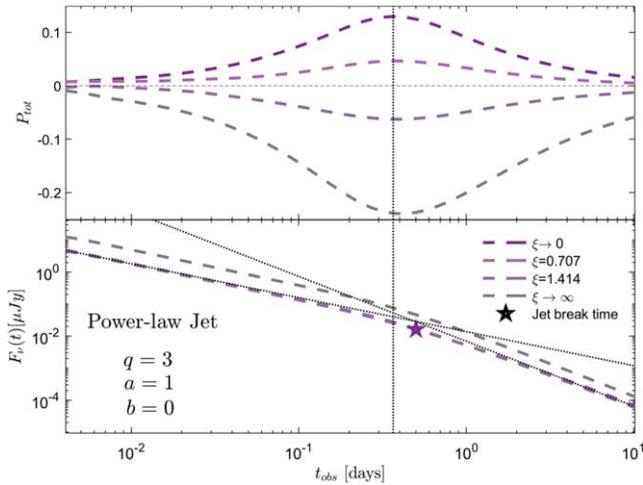

**Figure 4.** Observed polarization degree (upper panel) and flux (lower panel) at observer frequency $\nu = 10^{15}$ Hz (PLS G) with a flat distribution of initial Lorentz factor ($b=0$), smooth power-law energy profile with $a=1$, observed with $q=3$. The structure of the magnetic field changes with the value of $\xi$. The direction of the polarization vector changes by 90° once the $\xi$ parameter crosses 1. The polarization peak happens at about the same time for all curves and is close to the break time of the light curve (marked with a star). The light curves of the models corresponding to $\xi = 0.707, 1.414$, and $\xi \to 0$ converge to one another due to similar mean magnetic field magnitudes.

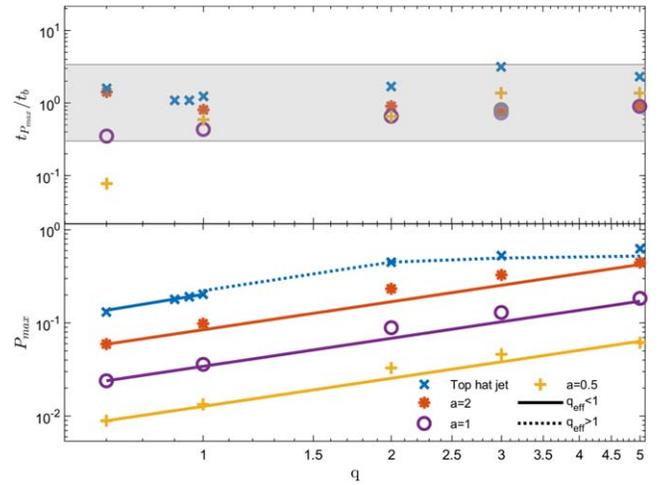

**Figure 5.** Upper panel: ratio between the polarization peak times $t_{P_{max}}$ and light-curve break times $t_b$ as a function of normalized viewing angle $q = \theta_{obs}/\theta_c$ for all jet structures considered in this work. We can see that these two times are within a factor of 3 from each other except for the extremely shallow structure of $a=0.5$, combined with a viewing angle of $q=0.7$ that forms a symmetrical system around the LOS and very late and slight jet break. Lower panel: maximal polarization degree $P_{max}$ as a function of the normalized viewing angle $q = \theta_{obs}/\theta_c$. For the top hat jet (blue × sign), the peak polarization for different $q$ values is described by two different functions for $q_{eff} \leqslant 1$ (solid lines) and $q_{eff} > 1$ (dotted lines). For the structured shallow jets, the maximal polarization degree seems to evolve as $P_{max} \propto q$ (solid lines), close to the $q<1$ behavior for the top hat jet.

We parameterize the 3D structure of the magnetic field following the formalism of R. Sari (1999) and R. Gill & J. Granot (2020), in which an initial field with a constant strength and an isotropic 3D orientations is stretched by a factor $\xi$ in the radial (shock normal) direction (for details see Appendix B). In this formalism, $\xi = 0$ corresponds to a field confined to the plane of the shock and $\xi \to \infty$ to a purely radial field.

Figure 4 shows the polarization curves and light curves from four structured jet models, all with power-law indexes of $a=1$ and $q=3$. The models differ by their $\xi$ values with $\xi = 0, 0.707, 1.414$, and $\infty$. In the first two models, where $\xi < 1$, the light from the perpendicular magnetic field component dominates, causing the polarization to have a positive sign as before. For the two models with $\xi > 1$, the polarization is mostly in the azimuthal direction and thus has a negative sign. It can also be seen that the maximal polarization is lower in the cases where $\xi$ is close to 1, corresponding to cases where $B'_{\parallel}$ and $B'_{\perp}$ are of the same order. The time evolution of the polarization degree is similar to what we discussed in Section 3.1.2. The absolute value of the polarization levels in the extreme magnetic field configurations corresponding to $\xi \to 0, \infty$ differ from one another. The $\xi \to \infty$ configuration features a uniform field, directed along the local velocity vector, and this asymmetry results in higher polarization degrees compared to those of the $\xi \to 0$ magnetic field configuration. The involvement of random components in the latter case leads to a rise in cancellations as contributions from radiating areas are summed to estimate the total polarization, thus reducing it.

### 3.3. Behavior of Peak Polarization

The temporal evolution of the observed polarization degree, demonstrated in Figures 2–4, shows that the value of the peak polarization degree, $P_{max}$, varies with changes in the viewing angle $q$, the jet wings energy power-law index $a$, and orientation of magnetic fields behind the shock, $\xi$. The dependence of $P_{max}$ on $q$ and $a$ can be explained in terms of a simple toy model for the polarization degree, inspired by the correlation between the time of peak polarization, $t_{P_{max}}$ and the jet break time $t_b$. The upper panel of Figure 5 demonstrates this correlation in jets with different structures and viewing angles. It shows that in almost all cases, the ratio between the two times varies between $1/3$ and 3. In the case of a jet with $a=0.5$ and $q=0.7$ we see a deviation from this relation, likely due to our inability to measure $t_b$ correctly. We evaluate $t_b$ by fitting two asymptotes to the afterglow light curve at early and late times and taking the intersection time between them, as shown in the lower panel of Figure 4. The outlier jet structure is extremely shallow and is observed from a relatively small effective viewing angle. As a result the jet break is very gradual and late, making it difficult to correctly fit an asymptote to the early-time afterglow light curve.

To model the polarization degree during the polarization peak time, we built a simplified toy model for the emission observed on the plane of the sky. We make the following simplifying assumptions:

1. We model a top hat jet, i.e., the emission comes only from the jet core.
2. The peak polarization time coincides with the time of the jet break, i.e., $t_{P_{max}} = t_b$.
3. The emission originates from a narrow arch across the core with an angular radius of $1/\Gamma$ from the LOS.

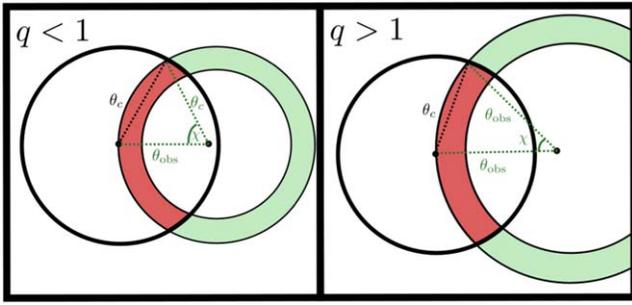

**Figure 6.** A toy model of the observed emission on the plane of the sky of a top hat jet during the polarization peak time, assuming that this time coincides with the jet break time. The left and right panels demonstrate cases of on-axis and off-axis jets, respectively. The black circle represents the angular size of the jet core ($\theta_c$), while the green ring covers the angular size of the observable region during the time of peak polarization with respect to an observer located at an angle $\theta_{\rm obs}$ from the jet axis. The red arches track the most luminous regions during the jet break time in both types of jets. The polarization degree is obtained by calculating the normalized $Q_\nu$ parameter of the light coming from the red arch, which scales as $Q_\nu/I_\nu \propto \sin(2\chi)/2\chi$. See the text for further details.

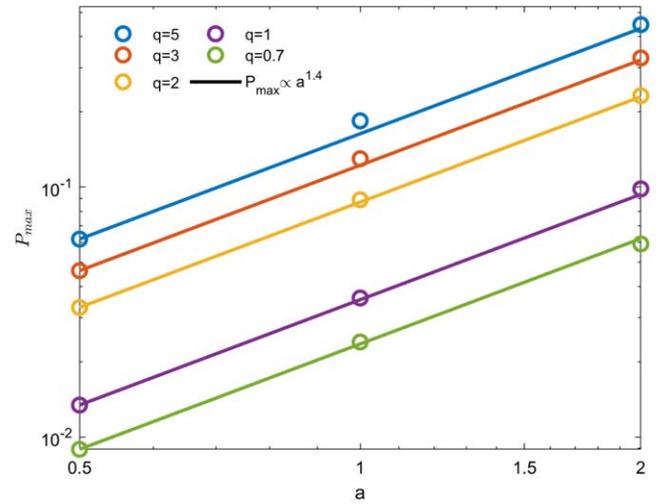

**Figure 7.** Maximal polarization degree $P_{\rm max}$ as a function of the energy profile of the jet, quantified as a declining power-law $a$. The maximal polarization degree seems to evolve as $P_{\rm max} \propto a^{1.4}$, which holds over all viewing angles.

Figure 6 illustrates our simple toy model for cases of on-axis (left) and off-axis (right) observers. The angular size of the jet core is marked by a black solid circle. The observer is located at an angle $\theta_{\rm obs}$ from the jet axis (the center of the black circle). The green ring centered around the observer's LOS marks the observable region at the time of $t_b$, assuming that at this time $\Gamma(\theta=0) = \theta_c^{-1}$ in the on-axis jet and $\Gamma(\theta=0) = \theta_{\rm obs}^{-1}$ in the off-axis jet. The red solid arch marks the region from which most of the observed emission originates. If the magnetic field is bound to the plane of the shock ($\xi \to 0$) then the light emitted from the red arch is polarized in a radial direction relative to the LOS on the plane of the sky. The total polarization degree in this case can be estimated by calculating the total $Q_\nu$ parameter of the light emitted by the arch, normalized by its total power, $I_\nu$:

$$P_{\rm max} = \frac{Q_\nu}{I_\nu} = \Pi_{\rm max} \frac{\int_{-\chi}^{\chi} \cos(2\phi)d\phi}{\int_{-\chi}^{\chi} d\phi} = \Pi_{\rm max} \frac{\sin(2\chi)}{2\chi}, \quad (21)$$

where $\phi$ is an azimuthal angle on the arch and $\chi$ is the angle of the arch at the edge of the jet core (see Figure 6). For on-axis jets ($q<1$) $\cos\chi = q/2$ and for off-axis jets ($q>1$) $\sin\chi = 1/2q$, as can be seen in Figure 6. Substituting these values in Equation (21) we get the value of $P_{\rm max}$ in the two asymptotic limits:

$$P_{\rm max} \simeq \Pi_{\rm max} \begin{cases} \frac{q}{\pi}\left(1 + \frac{q}{\pi}\right) & q \ll 1 \\ q\sin 1/q & q \gg 1. \end{cases} \quad (22)$$

Note that $P_{\rm max}$ passes smoothly through $q=1$. The two asymptotes intersect at $q=1.875$ at an approximated value of $P_{\rm max}$ that is ~15% larger than the value obtained from Equation (21); thus, in practice Equation (22) can provide a reasonable approximation for $P_{\rm max}$ at the entire range of viewing angles. The lower panel of Figure 5 shows the value of $P_{\rm max}$ in jets with different structures observed from various viewing angles. The color scheme is the same as in the upper panel. The dotted blue line represents a fit of Equation (21) to the data of the top hat jet, normalized by $N=0.8$. The data of the shallow jets ($a=0.5-2$), is fitted by the $q<1$ asymptote of

Equation (22). The shallow energy profiles in the wings of these jets increase the effective angular size of the core, leading to a smaller effective $q$ parameter $q_{\rm eff}$ than the naive value of $\theta_{\rm obs}/\theta_c$. Universal definitions of the effective core opening angle, based solely on the jet energy profile, may not fit all structured jets in general. For example, the definition given in T. Govreen-Segal & E. Nakar (2023) does not apply to shallow jets with $a<2$. We choose to quantify this $q_{\rm eff}$ parameter using the angular flux profile during the jet break time $t_b$, as the luminosity of a certain emitting region is the observational factor that determines whether or not it is a part of the core. We set the effective core opening angle at 10% of the angular luminosity cumulative distribution function peak. As a result, the semilinear fit of $P_{\rm max}(q)$ extends to larger values of $q$. We plan to explore this definition in the context of afterglow polarization in future work.

Figure 7 shows the change in $P_{\rm max}$ when we keep the $q$ parameter constant and change the value of $a$. Larger values of $a$ imply a steeper energy profile in the jet wings and a smaller effective core angular size. This leads to a larger effective value of the $q$ parameter and higher $P_{\rm max}$ for the same viewing angle, consistent with the above results. We find that $P_{\rm max} \propto a^{1.4}$ in all values of $q$ tested, implying that $q$ and $a$ are rather independent parameters. Stating this differently, the effective angular size of the core depends on the $a$ parameter and is the same for all observers viewing the jet from different angles. This relation also holds for both smooth power-law and broken power-law jet structures.

The above tests were performed assuming that the magnetic field is bound to the plane of the shock ($\xi \to 0$). In this situation the polarization degree in the most luminous region is close to $\Pi_{\rm max}$ and points along the radial direction from the observer. It is important to test how our results change when the value of $\xi$ varies. Figure 8 shows the change in $P_{\rm max}$ in three types of shallow jets observed at different viewing angles, as a function of $\xi$. All curves show qualitatively similar shapes that can be fitted by a hyperbolic tangent function:

$$P_{\rm max}(\xi) = A\tanh(-B\log_{10}\xi + C) - D, \quad (23)$$

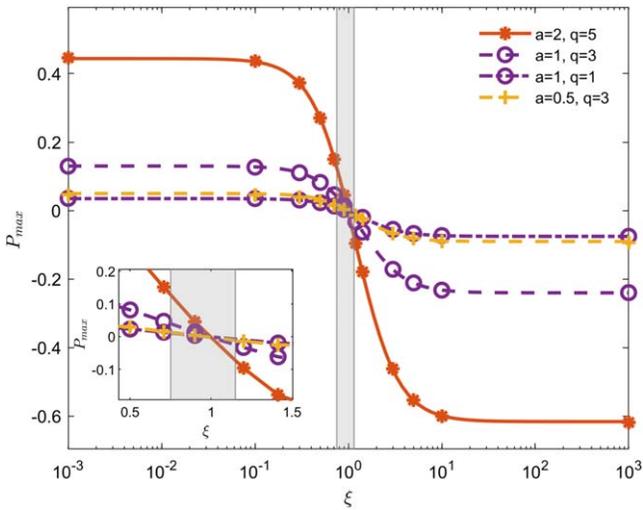

**Figure 8.** Maximal polarization degree $P_{\max}$ as a function of the magnetic field structure parameter $\xi$, under which synchrotron radiation is emitted. This relation is plotted for a few geometrical combinations and seems to agree with a hyperbolic tangent of $\log_{10}\xi$ for all values. We mark in shaded gray the limits on $\xi$ imposed by the upper limits on the polarization of GW170817, corresponding to $0.75 < \xi < 1.15$. In the insert we zoom in on the region around these limits. All curves change sign at $\xi = 1$. Detailed models can be found in Table 2.

where $(A - D)$, $-(A + D)$ are the asymptotic values of $P_{\max}$ at $\xi \to 0$ and $\xi \to \infty$ respectively, $B$ is the width of the transition region, and $C = \tanh^{-1}(D/A)$ ensures that $P_{\max}(\xi = 1) = 0$ as expected in a case of a perfectly isotropic magnetic field. Fittings of the function to the $P_{\max}(\xi)$ in cases with different $a$ and $q$ are given in Table 2.

Notice that Equation (23) is not symmetric about the x-axis, namely, $|P_{\max}(\xi \to 0)| - |P_{\max}(\xi \to \infty)| = -2D$ (instead of 0 in the symmetric case). The reason for that is that in the case of $\xi \to 0$ the polarization is obtained by summing contributions from regions with random field on the plane of the shock, whereas in the case of $\xi \to \infty$ the contributing field is ordered (radial), and produces a higher overall polarization. We also note that the asymmetry level $(D/A)$ decreases with a larger $a$. The reason is that the observed light in jets with steeper wings (larger $a$) is coming from narrower regions. When $\xi \to 0$ the polarization vector in each emitting zone in the ring will deviate less from this radial direction, giving rise an overall $P_{\max}$ that is closer to $P_{\max}(\xi \to \infty)$.[12] The immediate implication is that there is some dependency between $\xi$ and $a$; however, as long as we limit the results to $\xi \lesssim 1$ the dependency is weak, allowing us to obtain a simple analytic expression for $P_{\max}$ applicable for shallow jets:

$$P_{\max} = qa^{1.4}[0.055\tanh(-2.3\log_{10}\xi + 0.34) - 0.02]. \quad (24)$$

Following the approach of R. Gill & J. Granot (2020),[13] we use the limits on the radio polarization degree of GW170817 with $P_{\rm tot} < |12\%|$ at 244 days, along with its fitted power-law jet structure from R. Gill & J. Granot (2018), to constrain the typical value of the $\xi$ parameter in terms of our model. We find a nearly isotropic magnetic field structure agrees with the limits on GW170817 with $0.75 < \xi < 1.15$ (plotted gray area in Figure 8). Equation (24) is applicable to this range within the $\xi$ parameter space. Assuming a "single-zone" model, in which the same shock-generated field induces the multiwavelength synchrotron afterglow forward shock emission, allows us to generalize the constraints on $\xi$ from the radio to the optical regime. In the context of shallow jets, we can interpolate the results of Figure 4 using these limits and conclude the near isotropic magnetic field structure imposed by observations, would induce very modest polarization degrees upon shallow jet models, regardless of the off-axis viewing angle.

---

[12] The asymmetry level also slightly decreases for large $q$ since a larger fraction of the emission near the time of peak polarization is coming from an angle close to $1/\Gamma$ from the LOS, where $\hat{n}_{\rm obs}$ is close to the shock plane leading to a more ordered field in projection to the plane normal to $\hat{n}'_{\rm obs}$, which is closer to the ordered radial field case.
[13] Also derived independently by A. Corsi et al. (2018), E. Stringer & D. Lazzati (2020), and O. Teboul & N. J. Shaviv (2021).

Another parameter dependence we have not explored so far was the propagation of the relativistic shock into a nonuniform medium with $\rho \propto R^{-k}$ where $k > 0$. We plan to thoroughly explore this in future work.

## 4. GRB 221009A—A Shallow Jet?

GRB 221009A was a bright long GRB with an isotropic-equivalent energy $E_{\gamma,\rm iso} \approx 10^{55}$ erg (A. de Ugarte Postigo et al. 2022; B. O'Connor et al. 2023), the largest isotropic-equivalent gamma-ray energy measured to date (E. Burns et al. 2023). Detection of TeV photons from the early afterglow (Y. Huang et al. 2022) also suggests that the intrinsic energy of the event is quite high. Its measured low redshift of $z = 0.151$ (A. de Ugarte Postigo et al. 2022; D. B. Malesani et al. 2023), together with the high observed flux, allowed for detailed multiwavelength afterglow measurements (J. S. Bright et al. 2023; B. O'Connor et al. 2023; T. Laskar et al. 2023 M. A. Williams et al. 2023) as well as obtaining upper limits on the X-ray polarization of 13.8% placed 3.5 days post-burst (M. Negro et al. 2023).

Despite the plethora of observations, no classical jet break was detected in the afterglow light curve throughout the first 100 days after the GRB. Combined with the record-breaking high fluence and $E_{\gamma,\rm iso}$, the lack of a jet break poses a possible energy crisis for GRB progenitor models (B. O'Connor et al. 2023). One appealing solution is that the relativistic jet core is surrounded by energetic wings with a shallow angular energy profile. This assumption allows obtaining good fits to the afterglow light curves using a reasonable intrinsic GRB energy. However, degeneracies in the classical afterglow model imply that different jet parameters can fit the observed multiwavelength light curve (e.g., R. Gill & J. Granot 2023; B. O'Connor et al. 2023). R. Gill & J. Granot (2023) featured a flat cored jet with $\theta_c = 1.2°$ surrounded by wings with shallow power-law profiles in energy per solid angle and Lorentz factor, that scale as $\theta^{-0.8}$ and $\theta^{-0.3}$, respectively. The viewing angle corresponds to $q = 0.95$ and the surrounding medium has a wind-like power-law profile with $k = 2$. B. O'Connor et al. (2023) fitted a broken power-law jet with no core where both the inner and outer segments have shallow angular power-law profiles in the energy per solid angle, proportional to $\theta^{-0.75}$ and $\theta^{-1.15}$, respectively. The opening angle of the inner power-law segment is $3°$, the outer segment extends up to $23°$ and the jet spreads into a uniform density medium with $k = 0$. We refer the reader to Appendix F for a full description of the afterglow model parameters. Both models are able to explain the broadband spectrum with a combined contribution from the reverse and forward shocks, where the reverse shock dominates the radio emission during the first 100 days of observations. According to these models, the optical emission is reverse shock dominated up to 1 day and dominated by the forward shock at later times, while the X-ray emission is forward shock dominated throughout the entire evolution.

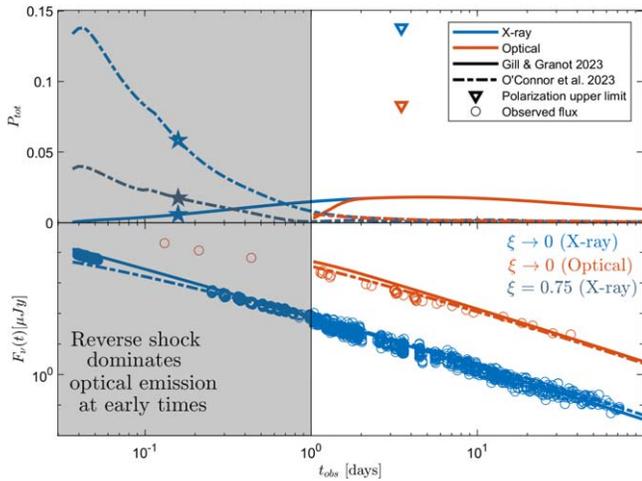

**Figure 9.** Upper panel: temporal evolution of the polarization degree for two suggested structures for GRB 221009A: R. Gill & J. Granot (2023) in solid lines and B. O'Connor et al. (2023) in dashed–dotted lines. The curves are computed in two bands: X-ray (blue) and optical (red) where the upper limits on linear polarization in both bands is presented with triangles in corresponding colors (M. Negro et al. 2023). The structure of the magnetic field is taken to be random on the plane of the emitting region ($\xi \to 0$). In dark blue dashed–dotted lines we plot the X-ray polarization curve for the B. O'Connor et al. (2023) model with $\xi = 0.75$ and find reduced polarization at early times. All proposed models agree with the upper limits on polarization taken at 3.5 days in both bands. We mark with stars the polarization level at ~3.8 hr post-burst. Lower panel: X-ray (blue) and optical (red) light curves for two of the proposed models for the structure of the jet of GRB 221009A: R. Gill & J. Granot (2023) in solid lines and B. O'Connor et al. (2023) in dashed–dotted lines. Both forward shock models fit the observed data (B. O'Connor et al. 2023; R. Gill & J. Granot 2023).

We reproduce the light curve fits with our numerical tool, according to the jet structures suggested by these models and present them in the lower panel of Figure 9. The solid lines present the R. Gill & J. Granot (2023) model and the dashed–dotted lines show the B. O'Connor et al. (2023) fit to the data. The empty circles represent the data points where the blue and red colors show X-ray and optical data and fits, respectively. Since our model evaluates emission from the forward shock, we present the optical fits only for times >1 day. The upper panel of Figure 9 shows the associated polarization curves with the same color mapping as in the bottom panel, assuming a random magnetic field confined to the plane of the shock ($\xi \to 0$). Additionally, we plot the X-ray polarization curve of B. O'Connor et al. (2023) with a more realistic value of $\xi = 0.75$ in dark blue dashed–dotted lines. The R. Gill & J. Granot (2023) model produces polarization levels $\to 0$ for $\xi = 0.75$. The open triangles mark the upper limits placed at 3.5 days post-burst by M. Negro et al. (2023) in X-ray and optical. It can be seen that all models predict polarization degrees well below the observed limits even for the most optimistic magnetic field configuration. With these current polarization measurements, we have no way of discriminating between the two suggested models. While the polarization degree is very low at 3.5 days for both models, the X-ray polarization differs greatly at early times.

At ~3.8 hr post-burst the difference between the polarization degrees predicted by the two models is ~5% in the most optimistic configuration and 1.75%, if using a more realistic $\xi = 0.75$ value (this time is marked with stars in the upper panel of Figure 9). The difference in the behavior of the polarization curves lies in the different angular energy profiles used at the jet cores. The model by R. Gill & J. Granot (2023) features a flat core profile with a narrow opening angle, similar to the profiles tested in this work, with the viewer located close to the core edge. The polarization curve in this case has a peak at a time close to the jet break time, when $\Gamma \simeq \theta_c^{-1}$, seen at about 1 day in the X-ray light curve. Note that the jet in this case expands into a wind-like medium with $k = 2$; thus, the polarization peak is smeared over more than a decade in time. In the model by B. O'Connor et al. (2023), the inner power-law segment features a distribution that peaks on the jet symmetry axis, while the viewing angle is very small. In this case the polarization peaks when the region near the jet symmetry axis is revealed as its Lorentz factor drops to $\Gamma = \theta_{\rm obs}^{-1} > \theta_c^{-1}$, and the slight jet break seen in the X-ray light curve is attributed to the broken power-law structure (B. O'Connor et al. 2023). Here, the jet expands into a uniform medium ($k = 0$), resulting in a narrower peak. The parameters used by the two models are presented in Appendix F. Had X-ray polarization been measured at earlier times, we should have been able to differentiate between the models via polarization modeling. In high-energy polarimetry, the instrumental ability to discern whether a signal is polarized or not can be estimated using the minimal detectable polarization (MDP$_{99}$). Low values of this quantity translate to more sensitive polarization measurements, which are able to distinguish low polarization degrees from the background. The extreme brightness of the X-ray afterglow of GRB 221009A at early times allows getting the MDP$_{99}$ of IXPE to low values, despite the short exposure times, since MDP$_{99} \propto$ [fluence]$^{-0.5}$ (M. C. Weisskopf et al. 2022). For example, if we take an optimistic IXPE repointing time of 1 hr post-burst, along with a minimal exposure time of 10 ks (~2.8 hr), the MDP$_{99} \leqslant 1.5\%$ at 3.8 hr post-burst. In addition to differentiating between the suggested jet structures, such a detection could have constrained the debated structure of the magnetic field behind the shock. In addition to the demonstrated theoretical benefits of a faster IXPE response time, Swift-XRT observations of GRB afterglow populations suggest that IXPE requires repointing within 6 hr of the trigger and exposure times on the order of several hundred kiloseconds to reliably detect X-ray polarization below 10% for at least one GRB annually (M. Negro & E. Burns 2024).[14]

## 5. Discussion

Traditional GRB models, assuming a top hat jet with uniform energy and Lorenz factor distributions, predict polarization degrees $\gtrsim 10\%$ at the time of the jet break when observing the jet on-axis ($0.5 \lesssim q = \theta_{\rm obs}/\theta_c < 1$; for smaller $q$ the polarization is lower) and even higher levels when observing off-axis with $q > 1$ (e.g., G. Ghisellini & D. Lazzati 1999; R. Sari 1999; J. Granot 2003; E. M. Rossi et al. 2004; L. Nava et al. 2016; G. Birenbaum & O. Bromberg 2021). However, the rapid decline in the observed flux with viewing angle in off-axis jets makes the latter case harder to observe (however, in some cases the off-axis jet emission can exceed that of an on-axis one, as shown in J. Granot et al. 2002). Meanwhile, most observed polarization levels remain at a level of 10% or below (e.g., S. Covino et al. 2004; I. A. Steele et al. 2009; T. Uehara et al. 2012; K. Wiersema et al. 2014), with a few outliers suspected to be associated with a reverse shock contribution, e.g., C. G. Mundell et al. (2013). Adding a jet structure to GRB afterglow models reduces the overall polarization and can relax the tension between theoretical models and observations.

---

[14] Link to the poster (https://aas243-aas.ipostersessions.com/default.aspx?s=D5-B6-7B-D0-4E-E0-39-40-78-94-7E-24-43-3A-D4-69&guestview=true).

We find that the peak polarization degree inversely depends on the amount of energy in the extended jet wings. We also found that the light contribution from the wings does not change the relation between the times of the jet break and peak polarization that was shown to exist in top hat jets (R. Sari 1999; J. Granot 2003; E. M. Rossi et al. 2004; G. Birenbaum & O. Bromberg 2021), it does however extend the break duration and reduce the change in the light-curve slope it induces. In the extreme case of shallow jets, the slope difference may be too small to be detected, resulting in a lack of an observed canonical jet break, together with extremely low polarization levels (e.g., B. O'Connor et al. 2023). The jet structure also delays the occurrence time of jet breaks seen by on-axis observers. This occurs as a result of an increase in the effective core angle due to emission from the jet wings. For off-axis observers, the energetic wings increase the visibility of the jet at early times while slightly changing the light-curve break time. This delay in the jet break time, following a decrease in the value of $a$ below a critical value corresponding to the transition to shallow jets, is also predicted from analytical arguments (see relations for $t_{\rm sh}$ in P. Beniamini et al. 2022).

A second factor that can reduce the polarization levels is the existence of a magnetic field component aligned with the shock normal in addition to the field on the shock plane, parameterized by $\xi$. For example, polarization measurements in the radio afterglow of GW170817, along with its light curve, indicate that the magnetic field is nearly isotropic with $0.75 < \xi < 1.15$ (adapted from the analysis of R. Gill & J. Granot 2020).

More generally, the observed polarization is affected by three main parameters: the jet structure, the magnetic field configuration, and the viewing angle. To map this apparent degeneracy we obtained an analytical expression connecting these three parameters to the peak polarization level, showing that $P_{\max} \propto q a^{1.4} F(\xi)$, where $q = \theta_{\rm obs}/\theta_c$ is the normalized viewing angle, $a$ is the power-law slope of the jet wing energy profile, and $F(\xi)$ is a shifted hyperbolic tangent (Equation (24)). Light-curve observations, together with polarization measurements during the jet break time, can help alleviate some of these degeneracies.

Polarization measurements at early times can provide an additional constraint on the jet structure. This can be especially important in the case of shallow jets where a jet break may not be detected and polarization levels are expected to be low, as we show for GRB 221009A. In this case, we showed that two different models of shallow jets, producing the same afterglow light curve, can have a significantly different level of polarization at early times (see Figure 9). Our polarization analysis focused on the X-ray band since it is not expected to be contaminated by light from the reverse shock and thus should reflect the magnetic field configuration generated by the forward shock alone. If future early-time X-ray polarization measurements are accompanied by the detection of optical or radio polarization as well, the combination of the two could be used to constrain the magnetization of the reverse shock, which probes the relic magnetic field in the outflow.

Low magnetization may be a common feature of shallow jets in general. We look into a sample of five other luminous events that share common features with GRB 221009A, such as long-lasting X-ray afterglows, lack of a canonical jet break in their light curves, and very high-energy emission, assembled by B. O'Connor et al. (2023). In some of these events there are early-time detections of optical and radio polarization, suspected to come from a reverse shock, that show low levels of polarization <8% (A. J. van der Horst et al. 2014; T. Laskar et al. 2019; N. Jordana-Mitjans et al. 2020; S. Dichiara et al. 2022; M. Arimoto et al. 2024). Numerical simulations show that hydrodynamic jets develop energetic wings with shallow energy profiles as a consequence of their passage through the confining medium, as opposed to their weakly magnetized counterparts, which form steeper profiles (O. Gottlieb et al. 2020, 2021; P. Beniamini et al. 2022). The polarization from the reverse shock in such jets is expected to be low due to the random orientation of magnetic fields behind it, making hydrodynamic jets suitable candidates for the observed shallow GRB jets.

Out of these five events, only two have measured polarization during the forward shock afterglow phase, showing levels <3% in the optical band (R. Itoh et al. 2013; A. J. van der Horst et al. 2014; M. Arimoto et al. 2024), consistent with the expected polarization from shallow jets. As stated above, early-time measurements of polarization in X-ray, optical, and radio bands can be used to constrain the magnetization in the ejecta shortly after it breaks out of the progenitor star. They may also help confirm the connection between the jet structure and the magnetization of the outflow and possibly differentiate between suggested models.

## 6. Conclusions

In this work we explore the effect of a power-law angular structure on GRB afterglow light curves and observed polarization. We find that increasing the amount of energy in the jet wings reduces the observed polarization and produces more moderate breaks in the afterglow light curves for both on-axis and off-axis jets. When the energy in the jet wings becomes comparable to the energy in the jet core, the jet break becomes extremely subtle and may not be observed. Such a distinctive feature is seen in energetic GRBs with long-lasting X-ray afterglows and very high-energy emission and may point to the involvement of an energetic extended structure. Analysis of our results allowed us to formulate an analytical relation between the system parameters and the peak polarization degree, supported by a physical toy model of the system. Finally, we employ our semi-analytical tool on the light curve fits for GRB 221009A and find an agreement with the upper limits placed on its afterglow polarization in both optical and X-ray. While this affirms the shallow jet models suggested for this event, it does not distinguish between them. Nevertheless, the polarization levels of both models considered for GRB 221009A differ greatly at early times. Possible IXPE detection could have allowed the desired distinction between them. The expected afterglow polarization degrees from shallow GRB jets are low, and if these are a common feature of energetic GRBs, future instrumentation should feature better sensitivity, along with faster response times.

## Acknowledgments

We thank Michela Negro for her insightful comments and help with understanding the capabilities of IXPE. In addition, we thank Brendan O'Connor and Geoffrey Ryan for their helpful discussions. We also thank the anonymous referee for perceptive suggestions. G.B. and O.B. are supported by ISF grants 1657/18 and 2067/22, a BSF grant 2018312, and an NSF-BSF grant 2020747. R. G. is supported by PAPIIT-2023 (IA105823) grant. P.B. is supported by a grant (No. 2020747) from the United States-Israel Binational Science Foundation (BSF), Jerusalem, Israel and by a grant (No. 1649/23) from the Israel Science Foundation.

## Data Availability

The data underlying this paper will be shared on reasonable request to the corresponding author.

## Appendix A
## Modeling the Forward Shock and the Magnetic Field Structure Behind It

The emitting region, generated by the structured relativistic shock that plows into the ISM, is modeled by a polar grid, which divides it into small angular sections $(\theta, \phi)$. These coordinates are centered around the jet axis. The polar angle $\theta$ is measured from the jet symmetry axis and the coordinate $\phi$ acts as an azimuthal angle. Our code uses two coordinate systems which are used to calculate physical qualities in the local and global frames. Local quantities on our 2D emitting region, which is meant to emulate a semispherical shock, are best described by the spherical coordinates $(\hat{r}, \hat{\theta}, \hat{\phi})$ (see panel (b) in Figure 10 for a detailed illustration). The radial direction is perpendicular to the face of the shock and parallel to the direction of the velocity at each cell. The $\hat{\theta}$ angular direction is measured from the radial direction and the $\hat{\phi}$ direction is denoted in the plane of the emitting surface. We demonstrate the use of the local spherical coordinate system by defining the direction of the magnetic field in the rest frame (panel (b) in Figure 10). Our model denotes the magnetic field as uniform in each cell with the following direction:

$$\hat{\boldsymbol{B}}' = \cos\theta_B \hat{\boldsymbol{r}} + \sin\theta_B \sin\phi_B \hat{\boldsymbol{\theta}} + \sin\theta_B \cos\phi_B \hat{\boldsymbol{\phi}}, \quad \text{(A1)}$$

where the angle $\theta_B$ is measured from the local radial direction and determines whether the magnetic field is confined to the plane of the shock or not. When $\theta_B = 90°$, the magnetic field is solely in the plane of the emitting surface $\in [\hat{\boldsymbol{\theta}} - \hat{\boldsymbol{\phi}}]$. The angle $\phi_B$ determines the orientation of the magnetic field on the plane of the emitting surface and it is measured from the $\hat{\phi}$ axis. A random magnetic field, with a small coherence scale, confined to the plane of the shock, would be denoted by randomizing the angle $\phi_B$ and setting $\theta_B = 90°$. This kind of structure can be explained by the generation of the field in the shock by plasma instabilities such as the two-stream Weibel instability (M. V. Medvedev & A. Loeb 1999; A. Achterberg & J. Wiersma 2007). In the case where $\theta_B = 0°$, the magnetic field is in the $\hat{r}$ direction and is considered radial. However, these two cases are extreme and the true structure of the magnetic field is probably a mixture of these two components. Since the magnetic field is uniform in each cell, the synchrotron radiation from each cell is maximally polarized.

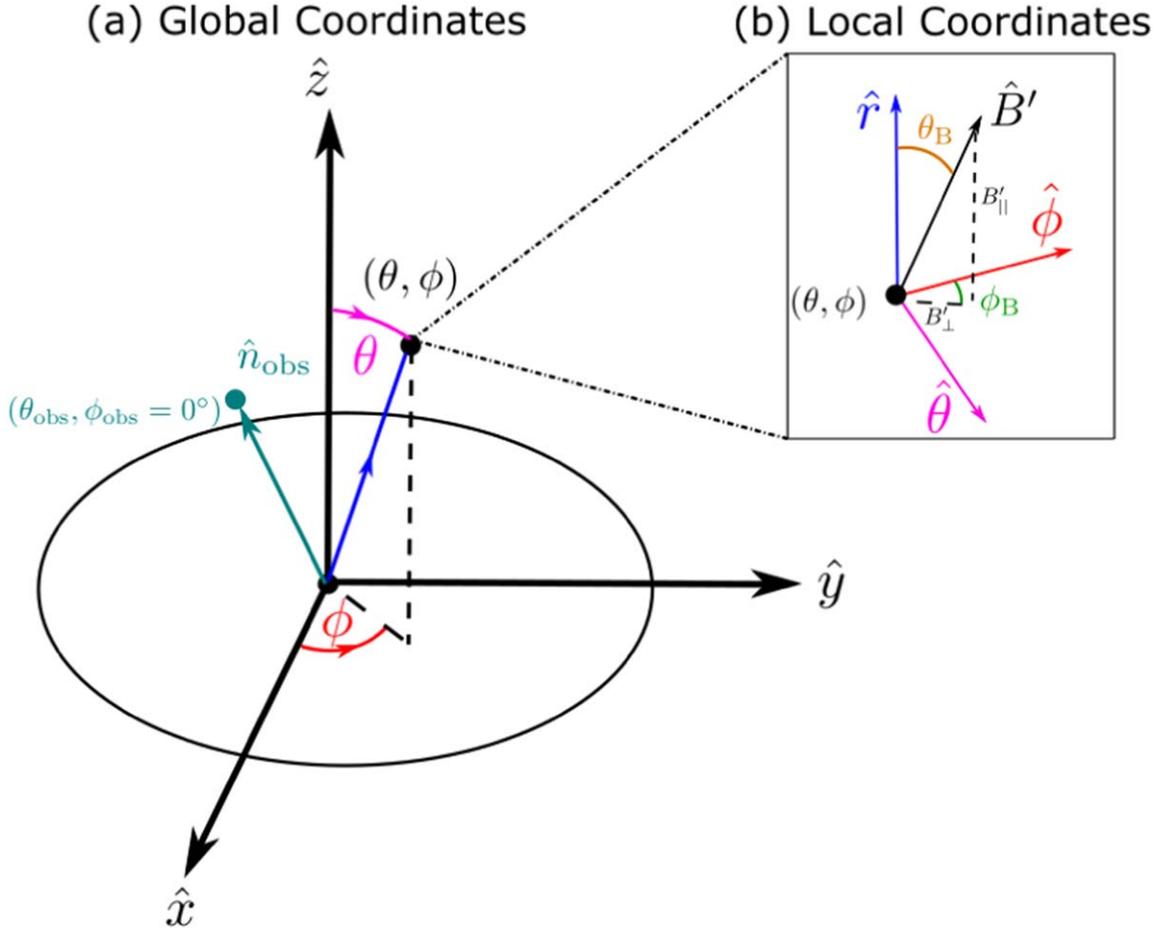

**Figure 10.** Illustration changed to right-handed coordinate systems. An illustration of the multiple coordinate systems used to evaluate polarization in structured GRB afterglows. (a) The global Cartesian coordinates define the symmetry axis of the jet as $\hat{z}$, the $\hat{x}$-axis connects the LOS (teal) and the symmetry axis for a choice of $\phi_{\rm obs} = 0°$ and the $\hat{y}$ axis is perpendicular to both, forming a right-handed coordinate system. (b) The local spherical coordinate system is denoted at each angular section $(\theta, \phi)$, where the radial direction (blue) is parallel to the velocity direction and the $(\hat{\theta}, \hat{\phi})$ directions (magenta and red, respectively) correspond to those denoted by the global coordinate system. We use this system, for example, to denote the magnetic field direction at each cell (see Equation (A1)). The angle $\theta_B$ (brown) is measured from the local radial direction and the angle $\phi_B$ (green) is measured from the local $\hat{\phi}$ direction.

Global quantities would be defined in terms of a Cartesian coordinate system and would be useful in determining quantities in the observer frame such as the direction to the observer (LOS). We assume the observer plane, from which emission reaches the observer, is perpendicular to the symmetry axis of the jet $\hat{z}$, which means it lies in the $x$–$y$ plane (see panel (a) in Figure 10). To later on evaluate the local emission quantities and polarization, we would also need to translate the global coordinates to the spherical local ones. The direction to the observer is a constant one and thus is defined by a constant vector in the global frame. For $\hat{n}'_{\text{obs}} = n_x \hat{x} + n_y \hat{y} + n_z \hat{z}$ and an observer located at the spherical angles [$\theta_{\text{obs}}$, $\phi_{\text{obs}}$], we can solve the following set of equations:

$$\begin{cases} n_x = \sin\theta_{\text{obs}}\cos\phi_{\text{obs}} = n_r \sin\theta\cos\phi + n_\theta \cos\theta\cos\phi - n_\phi \sin\phi \\ n_y = \sin\theta_{\text{obs}}\sin\phi_{\text{obs}} = n_r \sin\theta\sin\phi + n_\theta \cos\theta\sin\phi + n_\phi \cos\phi \\ n_z = \cos\theta_{\text{obs}} = n_r \cos\theta - n_\theta \sin\theta, \end{cases} \quad (A2)$$

where $\hat{n}'_{\text{obs}} = n_r \hat{r} + n_\theta \hat{\theta} + n_\phi \hat{\phi}$ is in terms of the local spherical coordinates. These turn out to be

$$\begin{cases} n_r = \sin\theta_{\text{obs}}\cos\phi_{\text{obs}}\sin\theta\cos\phi + \sin\theta_{\text{obs}}\sin\phi_{\text{obs}}\sin\theta\sin\phi + \cos\theta_{\text{obs}}\cos\theta \\ n_\theta = \sin\theta_{\text{obs}}\cos\phi_{\text{obs}}\cos\theta\cos\phi + \sin\theta_{\text{obs}}\sin\phi_{\text{obs}}\cos\theta\sin\phi - \cos\theta_{\text{obs}}\sin\theta \\ n_\phi = \sin\theta_{\text{obs}}\sin\phi_{\text{obs}}\cos\phi - \sin\theta_{\text{obs}}\cos\phi_{\text{obs}}\sin\phi. \end{cases} \quad (A3)$$

## Appendix B
## Magnetic Field Structure Parameterization

The downstream region, well behind the shock wave, can also contribute to the observed emission and may experience a stretching of the magnetic field component parallel to the bulk velocity direction, also denoted as the radial direction $\hat{r}$ (J. Granot 2003; R. Gill & J. Granot 2020). To account for this effect we adopt the formalism by R. Gill & J. Granot (2020), following R. Sari (1999), assuming a spherical isotropic field $\bar{B}'$ just behind the shock, which is first stretched only in the radial direction by a factor $\xi$ (while the field components in the plane normal to the radial direction remain unchanged), and then a global renormalization by a factor of $\sqrt{3/(2+\xi^2)}$ is applied to all field components in order for the field rms to remain the same as for the original isotropic field, $\bar{B}'$, namely,

$$\begin{aligned}\hat{B}' &= B'(\cos\theta_B \hat{r} + \sin\theta_B \sin\phi_B \hat{\theta} + \sin\theta_B \cos\phi_B \hat{\phi}) \\ &= \bar{B}' \sqrt{\frac{3}{2+\xi^2}}(\xi\cos\bar{\theta}_B \hat{r} + \sin\bar{\theta}_B \sin\bar{\phi}_B \hat{\theta} + \sin\bar{\theta}_B \cos\bar{\phi}_B \hat{\phi}),\end{aligned} \quad (B1)$$

where $\theta_B$ ($\bar{\theta}_B$) is the angle of the post-stretched (prestretched) proper field from the shock normal (as illustrated in Figure 10). The direction of the prestretched field is obtained from a uniformly distributed probability density per solid angle, namely, the angular parameters ($\bar{\mu}_B = \cos\bar{\theta}_B$, $\bar{\phi}_B$), are taken from uniform probability distributions ($-1 \leqslant \bar{\mu}_B \leqslant 1$, $0 \leqslant \bar{\phi}_B \leqslant 2\pi$) and then used to obtain the corresponding angular direction of the post-stretched field,

$$\mu_B = \cos\theta_B = \frac{\xi\bar{\mu}_B}{\sqrt{1+\bar{\mu}_B^2(\xi^2-1)}}, \qquad \phi_B = \bar{\phi}_B. \quad (B2)$$

This procedure is equivalent to obtaining $\mu_B$ from an angular probability distribution function[15]

$$f_\mu(\mu_B, \xi) = \frac{\frac{1}{2}\xi^2}{[\xi^2(1-\mu_B^2)+\mu_B^2]^{3/2}}. \quad (B3)$$

The generated values of ($\theta_B$, $\phi_B$) are then inserted into Equation (B1) to obtain the post-stretched field strength, which in contrast to the original isotropic field depends on $\mu_B = \cos\theta_B$ or $\bar{\mu}_B = \cos\bar{\theta}_B$,

$$\begin{aligned}S(\mu_B, \xi) &= \frac{B'}{\bar{B}'} = \sqrt{\frac{3\xi^2/(2+\xi^2)}{\xi^2(1-\mu_B^2)+\mu_B^2}}, \\ S(\bar{\mu}_B, \xi) &= \frac{B'}{\bar{B}'} = \sqrt{\frac{1+\bar{\mu}_B^2(\xi^2-1)}{\frac{1}{3}(2+\xi^2)}}.\end{aligned} \quad (B4)$$

In particular, $S(\bar{\mu}_B, 1) = 1$ (an isotropic field) as required, $S(\bar{\mu}_B, 0) = \sqrt{\frac{3}{2}}\sin\bar{\theta}_B$ (a field confined to the plane of the shock), and $S(\bar{\mu}_B, \xi \to \infty) = \sqrt{3}\,\bar{\mu}_B$ (a purely radial field).

---

[15] See R. Gill & J. Granot (2020), Appendix A for a detailed discussion.

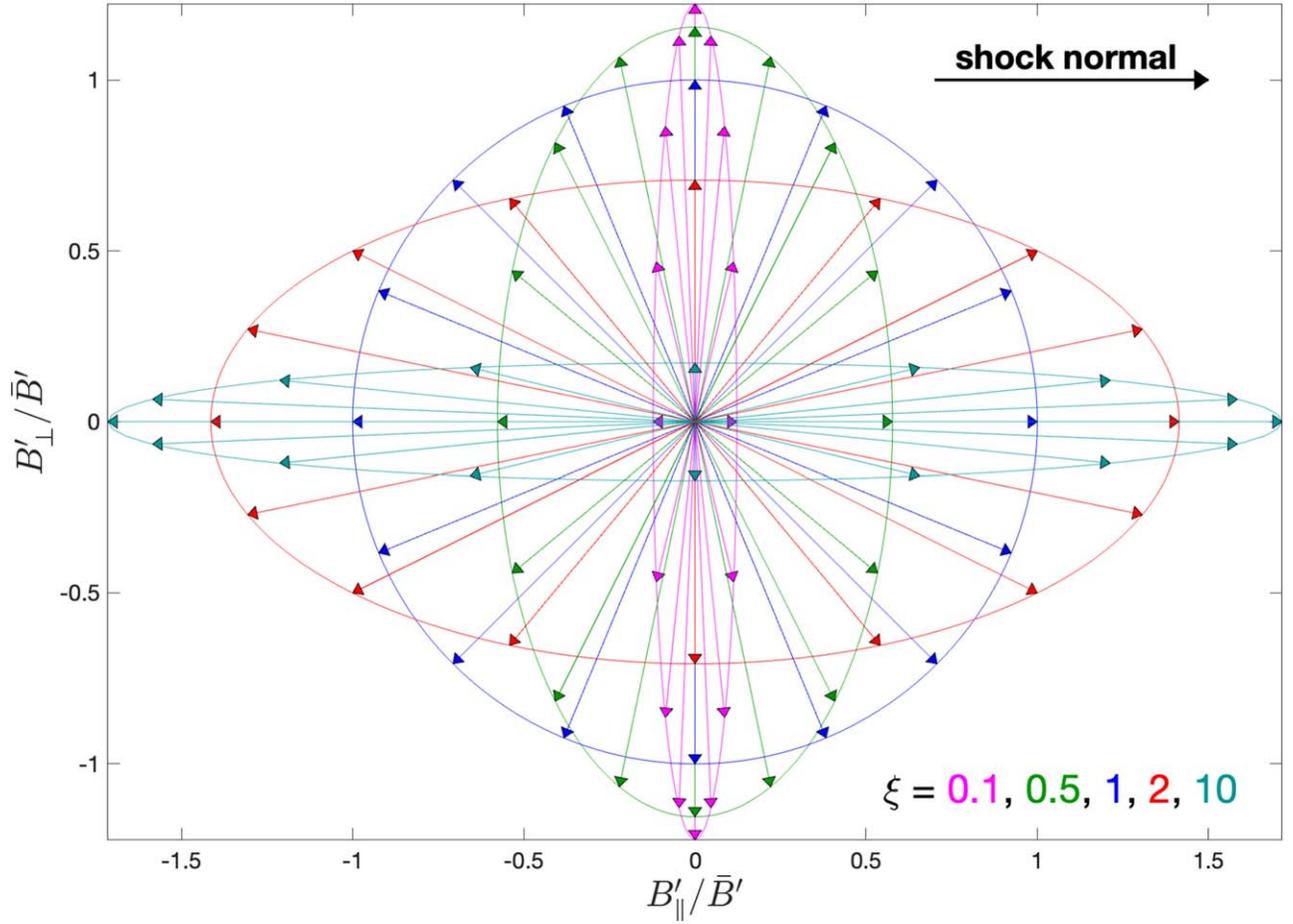

**Figure 11.** An illustration of our adopted local magnetic field structure for five different values of $\xi$.

Moreover, the rms field indeed remains unchanged,

$$\langle S^2 \rangle = \frac{1}{2} \int_{-1}^{1} d\bar{\mu}_B \, S^2(\bar{\mu}_B, \xi) = 1. \quad (\text{B5})$$

Figure 11 demonstrates our adopted local magnetic field structure for different values of $\xi$.

Our post-stretched magnetic field configuration in the case of $\xi \to 0$ is similar in terms of the field direction to that of previous works, which assumed a random magnetic field purely in the shock plane. This can be seen by substituting $\xi \to 0$ in Equation (B1), or equivalently by noting that the probability function $f_\mu(\mu_B, \xi)$ is narrowly peaked at $\mu_B = 0$ with an FWHM of the order $\Delta\mu_B \sim 2\xi$. However, our model is distinct from such works (in particular from J. Granot & A. Konigl 2003 who introduced the parameter $b$ with $b = 0$ used for this case), in terms of the field strength. While most previous works assumed a constant field strength, we effectively assume a random field strength with a cumulative probability distribution of $P(>S) = 1 - P(<S) = \sqrt{1 - \frac{2}{3}S^2}$ where the normalized post-stretch magnetic field strength, $S = B'/\bar{B}'$, assumes values in the range $0 \leqslant S \leqslant \sqrt{3/2}$. This difference can lead to some modest quantitative differences in the emission properties, while the qualitative emission and polarization properties remain very similar.

It is also worth noting the main difference between our model and that of R. Gill & J. Granot (2020), which also uses the parameter $\xi$ following R. Sari (1999), namely, the integration region. While we integrate over the 2D contribution to the observed emission from the immediate downstream region behind the shock wave, R. Gill & J. Granot (2020) integrate over the entire volume behind the shock. Moreover, they assume some value $\xi_f$ of the parameter $\xi$ just behind the shock, and follow its evolution behind the shock locally following the R. D. Blandford & C. F. McKee (1976) solution (assuming flux freezing). Since the radial stretching of each fluid element is larger than its tangential stretching, $\xi$ grows with the distance behind the shock. For this reason, the effective value of $\xi$ when accounting for the emission from the whole shocked region, $\xi_{\text{eff}}$, is somewhat larger than its value just behind the shock, $\xi_f$ (see Figure 4 of R. Gill & J. Granot 2020). Even though these are different assumptions, the polarization curves of both models are relatively close for $\xi \to 0$, as can be seen in Figure 5 of R. Gill & J. Granot (2020). Similar final results for $\xi > 1$ values would correspond to differing $\xi_f$ parameters in terms of R. Gill & J. Granot (2020) and this current work. A good example of this difference can be given by a choice of $\xi = 1$, which produces a completely random magnetic field, leading to a zero polarization degree, in terms of our model. On the other hand, the R. Gill & J. Granot (2020) model manages to produce polarization for $\xi_f = 1$, due to additional stretching of the magnetic field in regions well behind the shock, which leads to $\xi_{\text{eff}} > \xi_f = 1$.

## Appendix C
## Transforming between the Observer Frame and the Rest Frame

Now that we have the direction to the observer in the observer frame expressed in terms of the local spherical coordinates (see Equation (A3)), all that is left to do is transform it to the rest frame. This is done in the local frame as photons from each cell are boosted differently according to their angular inclination from the LOS, energy, and Lorentz factor. The transformation is done via rotation on the LOS $\hat{\boldsymbol{n}}'_{\rm obs}$ by an angle

$$\eta = \cos^{-1}\left(\frac{\tilde{\mu} - \beta}{1 - \beta\tilde{\mu}}\right) - \cos^{-1}(\tilde{\mu}), \quad (C1)$$

which is done about the axis

$$\hat{\boldsymbol{\eta}} = \hat{\boldsymbol{\beta}} \times \hat{\boldsymbol{n}}'_{\rm obs}, \quad (C2)$$

and $\tilde{\mu} = \hat{\boldsymbol{\beta}} \cdot \hat{\boldsymbol{n}}'_{\rm obs}$. This rotation is then applied in the direction of the photon using a rotation matrix (G. Birenbaum & O. Bromberg 2021):

$$\hat{\boldsymbol{n}}'_{\rm obs} = \boldsymbol{R}(\eta, \hat{\boldsymbol{\eta}})\hat{\boldsymbol{n}}'_{\rm obs}. \quad (C3)$$

This method will also allow us to transform quantities from the rest frame to the observer frame by using the inverse of the rotation matrix $\boldsymbol{R}(\eta, \hat{\boldsymbol{\eta}})$.

## Appendix D
## The Polarization Vector in the Observer Frame

The observer plane, upon which we evaluate polarization, is perpendicular to the symmetry axis of the jet $\hat{\boldsymbol{z}}$, which means it lies in the $\hat{\boldsymbol{x}}$–$\hat{\boldsymbol{y}}$ plane. Therefore, the correct and general way to represent the local polarization vector in the observer frame would be in terms of the global $\hat{\boldsymbol{x}}$–$\hat{\boldsymbol{y}}$ plane. We start by calculating the direction of the polarization vector in each cell. This is done in the rest frame in terms of the local coordinates by using the expressions in Equations (A1) and (C3):

$$\hat{\boldsymbol{P}}' = \frac{\hat{\boldsymbol{n}}'_{\rm obs} \times \hat{\boldsymbol{B}}'}{|\hat{\boldsymbol{n}}'_{\rm obs} \times \hat{\boldsymbol{B}}'|} \equiv P'_r\hat{\boldsymbol{r}} + P'_\theta\hat{\boldsymbol{\theta}} + P'_\phi\hat{\boldsymbol{\phi}}. \quad (D1)$$

We would then transform the rest frame polarization vector to the observer frame using the inverse of the rotation matrix presented in Equation (C3). This expression is then translated to the global coordinate system, similar to the way done in Equations (A2) and (A3), which gives an observer frame polarization unit vector of $\hat{\boldsymbol{P}} = P_x\hat{\boldsymbol{x}} + P_y\hat{\boldsymbol{y}} + P_z\hat{\boldsymbol{z}}$. Since the observed polarization vector lies in the $\hat{\boldsymbol{x}}$–$\hat{\boldsymbol{y}}$ plane, we will then use these vector components to evaluate the observed local polarization angle $\phi_{p,0}$ (see Equation (17)).

## Appendix E
## Functional Dependence of Peak Polarization on $\xi$

The analytical expression that relates the peak polarization degree to the magnetic field structure parameter $\xi$ has a hyperbolic tangent functional form (Equation (23)). In Table 2 we present the functional fits plotted in Figure 8 for different parameter combinations.

Table 2
Analytical Functional Dependence of the Polarization Peak Height for the Different Models Considered in Figure 8

| Parameters | $P_{\max}(\xi)$ Functional Dependence |
|---|---|
| $a = 2, q = 5$ | $0.53 \tanh(-2.2 \log_{10} \xi + 0.15) - 0.086$ |
| $a = 1, q = 3$ | $0.185 \tanh(-2.2 \log_{10} \xi + 0.3) - 0.055$ |
| $a = 1, q = 1$ | $0.055 \tanh(-2.3 \log_{10} \xi + 0.34) - 0.02$ |
| $a = 0.5, q = 3$ | $0.07 \tanh(-2.2 \log_{10} \xi + 0.3) - 0.02$ |

## Appendix F
## Model Parameters for GRB 221009A

To reproduce the light-curve fits of the two geometrical models considered in this work for the structure of GRB 221009A, we use their suggested afterglow parameters. The multiband light-curve fit in B. O'Connor et al. (2023) is produced using the public semi-analytical code afterglowpy, which allows for different energy angular profiles within the jet while fitting the observed flux (G. Ryan et al. 2020). The forward shock in this model plunges into a constant density external medium with number density $n_{\rm ISM} = 1$ [cm$^{-3}$]. afterglowpy was calibrated with the top hat jets models of BoxFit, which capture the complex dynamics of the system using a collection of 2D numerical simulations and bases light-curve calculations off these models (H. van Eerten et al. 2012). The JetFit package contains a more complex suite of 2D simulations that allows for the "boosted fireball" model and was used to fit the multiwavelength afterglow of GW170817 (Y. Wu & A. MacFadyen 2018). While our semi-analytical afterglow model reproduced the light-curve shape for the jet structure suggested for GRB 221009A in B. O'Connor et al. (2023), its normalization was incompatible due to different realizations of the radiation model. To account for that while preserving the jet structure we wish to probe, we modify their suggested afterglow model. We do so by setting $E_{\rm k,iso} = 10^{55}$ erg on the LOS and normalizing the energy distribution accordingly. The modified afterglow parameters required to reproduce the fit are $\xi_e = 0.0046$, $\epsilon_e = 0.0126$, and $\epsilon_b = 8.5 \times 10^{-6}$. While not directly dependent on simulation output for the dynamics of the jets, the strength of semi-analytical models for light-curve fittings, such as afterglowpy and our own, is their flexibility in jet structure, which allows for 3D-simulation-motivated input. These simulations are thought to better capture the jet structure formation through mixing of the inner light jet material with the heavier cocoon that surrounds it (O. Gottlieb et al. 2021). A version of these structures is adopted by R. Gill & J. Granot (2023) in order to explain the multiwavelength afterglow of GRB 221009A with their semi-analytical fitting tool, while accounting for a wind-like external medium. A similar formalism is adopted in this work and the light-curve fit is reproduced as is, along with accompanying polarization curves. The afterglow parameters for both models used in this work are presented in Table 3.

**Table 3**
Afterglow Parameters for the R. Gill & J. Granot (2023) Model and Modified Parameters for the B. O'Connor et al. (2023) Model

| Parameter | Afterglow Model Parameters GRB 221009A | |
|---|---|---|
| | R. Gill & J. Granot (2023) | B. O'Connor et al. (2023) |
| $\theta_c$ | 1.2° | 3° |
| $\theta_{\rm obs}$ | 1.14° | 0.57° |
| $\Gamma_c$ | 300 | 500 |
| $a$ (PL index of energy structure) | 0.8 | 0.75, 1.15 |
| $b$ (PL index of Lorentz factor structure) | 0.3 | 0 |
| $E_c$ [erg] | $2 \times 10^{55}$ | $10^{55}$ on the LOS |
| $n_{\rm ISM}$ [cm$^{-3}$] | 0.1 | 1 |
| Density profile ($n \propto r^{-k}$) | $k = 2$ | $k = 0$ |
| $p$ | 2.4 | 2.25 |
| $\epsilon_e$ | 0.01 | 0.0126 |
| $\epsilon_B$ | $10^{-4}$ | $8.5 \times 10^{-6}$ |
| $\chi_e$ | 0.01 | 0.0046 |
| $A_*$ | 0.33 [cm$^{-1}$] | 1 [cm$^{-3}$] |

## ORCID iDs

Ramandeep Gill  https://orcid.org/0000-0003-0516-2968
Paz Beniamini  https://orcid.org/0000-0001-7833-1043
Jonathan Granot  https://orcid.org/0000-0001-8530-8941